\documentclass[%
 aps, twocolumn,
showpacs,
 amsmath,amssymb,
prb,
]{revtex4}

\usepackage{graphicx}% Include figure files
\usepackage{dcolumn}% Align table columns on decimal point
\usepackage{bm}% bold math

\usepackage{color}

\begin{document}

%\preprint{APS/123-QED}

\title{Finite Size Effect in the Quantum Anomalous Hall system}% Force line breaks with \\
%\thanks{Footnote to title of article.}

\author{Hua-Hua Fu}
\email{hhfu@mail.hust.edu.cn}
\affiliation{College of Physics and Wuhan National High Magnetic field center, Huazhong University of Science and Technology, Wuhan 430074, China.}%Lines break automatically or can be forced with \\

\author{Jing-Tao L\"{u}}%
\affiliation{College of Physics and Wuhan National High Magnetic
field center, Huazhong University of Science and Technology, Wuhan
430074, China.}%

\author{Jin-Hua Gao}%
\email{jinhua@mail.hust.edu.cn}
\affiliation{College of Physics and
Wuhan National High Magnetic field center, Huazhong University of
Science and Technology, Wuhan
430074, China.}%

\date{\today}% It is always \today, today,

\begin{abstract}
We theoretically investigate the finite size effect in quantum
anomalous Hall (QAH) system. Using Mn-doped HgTe quantum well as an
example, we demonstrate that the coupling between the edge states is
spin dependent, and is related not only to the distance between the
edges but also to the doping concentration. Thus, with proper tuning
of the two, we can get four kinds of transport regimes: quantum spin
Hall regime, QAH regime, edge conducting regime, and normal
insulator regime. These transport regimes have distinguishing edge conducting properties while the bulk is insulting.
Our results give a general picture of the
finite size effect in QAH system, and are important for the transport experiments in QAH nanomaterials as well as  future
device applications.

%\begin{description}

%\item[PACS number(s)]
%73.43.-f, 72.25.Dc, 85.75.-d, 75.50.Pp

%\end{description}
\end{abstract}

\maketitle

Quantum anomalous Hall (QAH) state is a new state of matter of two
dimensional (2D) insulator, which has a quantized Hall conductivity
without Landau level.\cite{1,2,3,4} Energy band of the QAH state is
topological nontrivial and can be characterized by the first Chern
number, similar to the quantum Hall state. \cite{5,6}
Different from
the normal quantum hall state, the QAH state does not need an external magnetic field to break
the time reversal symmetry (TRS).\cite{7,8,9,10}
Haldane proposed
the first model of QAH state on honeycomb lattice in 1988. \cite{11}
Stimulated by the discovery of topological insulator (TI), \cite{12}
there has been extensive effort to find QAH state in realistic
materials, based on theoretical proposals, e.g., mercury-based
quantum wells, \cite{5} disorder induced Anderson insulators,
\cite{6} graphene system, \cite{13,14} silience,\cite{15} and
magnetic topological insulators.  \cite{10} This has led to its
recent experimental observation in magnetic TI of Cr-doped
$(\textrm{Bi},\textrm{Sb})_2 \textrm{Te}_3$. \cite{16} Numerous
exotic properties of the QAH state are yet to be explored, which are
not only of fundamental importance but also have potential
applications.

In this work, we theoretically investigate the QAH system in a
finite stripe geometry, where the finite size effect can
influence its transport property. The same effect in quantum spin Hall (QSH)
system, \cite{17} as well as in the 3D topological insulators,
\cite{18,19} has been studied in details in last few years. A key
finding is that, when the sample is narrow enough,
the coupling of edge states will open an energy gap,
and the quantized conductance of the edge states disappears. Thus,
there exists a critical width in QSH system, below which the gapless edge states
are destroyed.  The finite size effect is crucial for the device application, since it determines the transport property of small device .
Here, we illustrate that the finite size effect in the QAH system is more
complex than that in QSH system.  Due to the magnetic doping, the coupling between edge states becomes spin dependent, and is also related to the doping concentration.
Given the width of ribbon and the doping concentration, we can distinguish four kinds of transport regimes in a QAH ribbon. While the bulk is insulating in all these transport regimes, they have different edge conducting behaviors. Because that the edge conducting channel is the key feature of  topological nontrivial system and is the base of many novel quantum electronic devices, our results may be important for the QAH state in nanomaterials as well as the device application of the QAH system.

The system we study is a Mn-doped HgTe quantum well (QW). Without Mn doping, its
electronic states can be described by an effective four band model.
QSH effect occurs at certain well thickness. \cite{20} Note that
HgTe QW is the first experimentaly confirmed QSH system, \cite{21} where the
finite size effect is well understood and is much more obvious than that in
other TI materials. \cite{17} Therefore, it is an ideal example to start with,
and it will be easier to compare the finite size effect of QAH with that of
QSH system. But we emphasize here that our results are not just limited to the
HgMnTe system. They actually offer a general picture for the finite size effect of the QAH state in
magnetic doped TI systems.

As a starting point, we introduce the effective Hamiltonian of
the HgTe QW. The HgTe QW has an inverted band structure, where the
$p$-type $\Gamma_{8}$ has higher energy compared to the $s$-type
$\Gamma_{6}$ band at the $\Gamma$ point. The QSH phase appears as
the QW thickness crosses over a critical value. \cite{20} The HgTe
QW can be described by an effective four-band model\cite{5}
\begin{eqnarray}\label{eq:h}
H_{0}(k)=\left(
\begin{array}{cc}
h_{+}(\mathbf{k})&0\\
0&h_{-}(\mathbf{k})\\
\end{array}
\right),
\end{eqnarray}
with the basis of \{$|E_1,\frac{1}{2}\rangle$,
$|H_1,\frac{3}{2}\rangle$, $|E_1,-\frac{1}{2}\rangle$,
$|H_1,-\frac{3}{2}\rangle$\}.
$h_{+}(\mathbf{k})={M_1}(\mathbf{k}_x\sigma_x-\mathbf{k}_y\sigma_y)+(M_{0}+M_{2}\mathbf{k}\cdot\mathbf{k})\sigma_z+\epsilon_\mathbf{k}$
where $\sigma_{x,y,z}$ are Pauli matrix.
$h_{-}(\mathbf{k})=h^{\ast}_{+}(\mathbf{-k})$ is required by TRS,
and $\epsilon_\mathbf{k}=C_{0}+C_{2}k^2$. The parameters $M_{0}$,
$C_{0}$, $C_{2}$, $M_{1}$ and $M_2$ depend on the thickness of the quantum well and the materials details
, which can be obtained in Refs. 22 and 23. $h_{\pm}(k)$ is actually
equivalent to two dimensional Dirac model, which has a quantized
Hall conductance $\pm e^2/h$. Thus, the net Hall conductance of the
HgTe QW is zero but the spin Hall conductance is nonzero. The QSH
state can be viewed as two copies of QAH states with opposite Hall
conductance. For numerical calculation, the tight-binding
representation of the four-band Hamiltonian on square lattice is
used. \cite{24,25,26}

To realize the QAH phase, magnetic doping is needed to break the TRS between
the two spin blocks. When one spin block is in normal insulting regime, but the
other is topological nontrivial, QAH state appears, resulting in nonzero net
Hall conductance. The spin splitting induced by the magnetization of the Mn
atoms is described by a phenomenological term
\begin{eqnarray}
H_{s}=\left(
\begin{array}{cccc}
G_{E}&0&0&0\\
0&G_{H}&0&0\\
0&0&-G_{E}&0\\
0&0&0&-G_{H}\\
\end{array}
\right),
\end{eqnarray}
e.g., 2$G_{E}$ for the $|E_{1}, \pm \frac{1}{2} \rangle$ hand and
2$G_{H}$ for the $|H_{1}, \pm \frac{3}{2} \rangle$ band. Thus the
energy gap is given by $\Delta{E_{\uparrow}}=2M_{0}+G_{E}-G_{H}$ for
the up-spin block, $\Delta{E_{\downarrow}}=2M_{0}-G_{E}+G_{H}$ for
the down-spin block. The QAH phase in the
Hg$_{1-\textbf{y}}$Mn$_{\textbf{y}}$Te QW has been discussed in
detail in Ref. 5, and here we just give a short summary. In order to
achieve the QAH phase in the Hg$_{1-\textbf{y}}$Mn$_{\textbf{y}}$Te
QW, a key relation $G_{E}G_{H}<0$ is required.
%From a standard perturbative
%treatment of the eight-band Kane model, \cite{18,24} one can find that
The expressions of $G_E$ and $G_H$ are $G_{E}=-(3AF_{1}+BF_4)$ and
$G_{H}=-3B$, where $F_1$ ($F_4$) is the amplitude of $\Gamma_6$
($\Gamma_8$) component in the states $|E_{1}, \pm \frac{1}{2}
\rangle$,  $A=\frac{1}{6}N_{0}\alpha{y}\langle{S}\rangle$ and
$B=\frac{1}{6}N_{0}\beta{y}\langle{S}\rangle$. \cite{27,28} $N_{0}$
is the number of unit cells per unit volume. $\alpha$ and $\beta$
describe $sp-d$ exchange coupling strength for the \emph{s}-band and
the \emph{p}-band electron. For
Hg$_{1-\mathbf{y}}$Mn$_\mathbf{y}$Te, the parameters are $F_1=0.57$,
$F_4=0.43$,  $N_{0}\alpha=$0.4 eV and $N_{0}\beta=$-0.6 eV. The Mn
doping is described by two parameters, i.e. doping
concentration $y$ and the single Mn atom spin polarization out of QW plane
$\langle{S}\rangle$. With proper doping parameters, e.g. $y$=0.02
and $\langle{S}\rangle = 2$, we can get QAH phase in this system.

\begin{figure}
\includegraphics[width=3.4in]{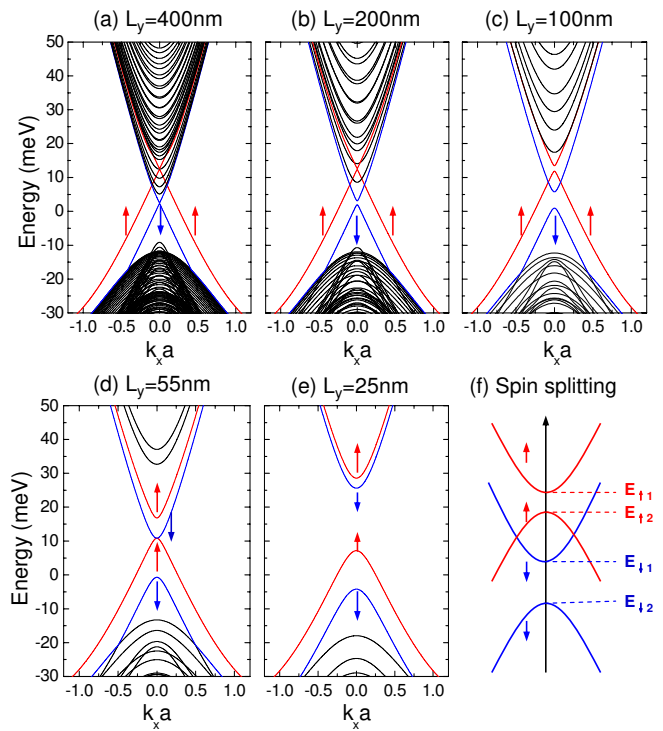}% Here is how to import EPS art
\caption{\label{fig:wide}(Color online) The band structure and edge
states upon decreasing the Hg$_{(1-y)}$M$_y$Te width $L_y$ for the
Mn dopping $y=0.02$ and the Mn spin polarization
$\langle{S}\rangle$=1 in the figures (a)-(e).
In the figure (f), we illustrate the definition of  $E_{1\uparrow}$ ($E_{1\downarrow}$) and
$E_{2\uparrow}$ ($E_{2\downarrow}$), which  are the energies of two
spin-up (spin-down) edge bands at the zero momentum points. The definition is useful for following discusions}
\end{figure}

Our goal is to explore the finite size effect in the QAH system, so
we investigate the Hg$_{1-\mathbf{y}}$Mn$_\mathbf{y}$Te QW in a
stripe geometry with width $L_y$ ($x$ direction is infinite). The
key feature of finite size effect is the gap opening in the energy
dispersion of the edge states, resulted from their mutual coupling.
In Fig. 1, we plot the energy spectrum of the strip with sample
width $L_y$ from 400 nm to 25 nm ($y=0.02$ and
$\langle{S}\rangle=1$), where red (blue) line indicates the
dispersion of the up (down) spin edge states. In Fig. 1(a), the sample width is 400 nm. The
$L_y$ is large enough, so that there is no coupling between edge
states. We can see that the dispersion of edge states is gapless for
both up and down spin. The sample is in the QSH regime, the charge
conductance is $2e^2/h$ and robust against disorder. \cite{29} Note that here the QSH state is TRS broken due to the magnetic doping\cite{30}.
Decreasing the width to 200 nm, see Fig. 1 (b), the coupling between
the down spin edge states is strong enough to open a measurable gap,
while the up spin states remain gapless. As a result, charge
conduction through the down spin edge channels is no longer
topological protected, and backscattering by impurity becomes
possible. Hence, with some disorder, the charge conductance in this
regime should range from $2e^2/h$ to $e^2/h$. An typical case is
that, when the Fermi level is in the energy gap of down spin states,
the charge conductance becomes $e^2/h$ and the transport property is
now similar as that of the QAH state. So we call this transport
regime as QAH regime. But it should be emphasized that  the QAH
transport regime here results from the finite size effect, different from the
normal conception of QAH state. In Fig. 1(c), when the sample width
$L_y$ is decreased to 100 nm, all the edge states become gapped.
Meanwhile, due to the spin splitting, near the $k_x=0$ point, the
down band of spin up is higher than the up band of spin down. No
matter where the Fermi level is (still in the bulk gap), there are
always conducting edge channels near the Fermi level. This is the
edge conducting regime, in which conduction through the edge can be
completely killed by disorder. The conductance can range from
$2e^2/h$ to 0. Decreasing $L_y$ further, the energy gaps increase
and we get a normal insulator, see Fig. 1(d) and (e). From above results, we
see that the coupling between edge states in the QAH system is spin
dependent. This is not surprising, because the two spin blocks in
the Hamiltonian [Eq.~\eqref{eq:h}] are no longer related by TRS, due
to magnetic doping. A direct consequence is that there are four
kinds of transport regime: QSH regime, QAH regime, edge conducting
(EC) regime and normal insulator (NI) regime.

\begin{figure}
\includegraphics[width=3.4in]{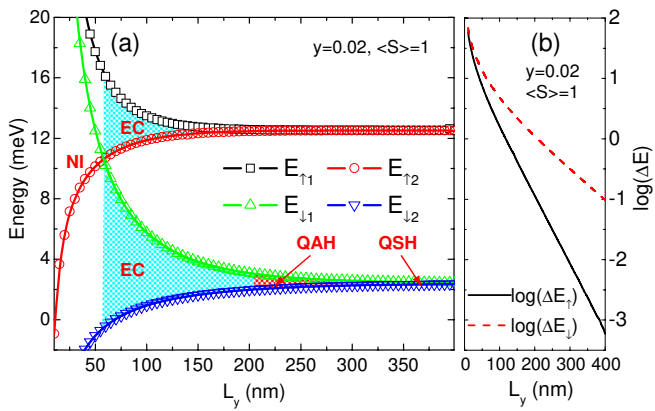}% Here is how to import EPS art
\caption{\label{fig:wide}(Color online) (a) The band energies
$E_{1\uparrow}$, $E_{2\uparrow}$, $E_{1\downarrow}$ and
$E_{2\downarrow}$ as a function of QW width $L_y$, with the Mn spin polarization $\langle{S}\rangle=1$ and 
the doping $y=0.02$.  The transport regimes are shown as well. (b) The energy
gaps $\mathbf{log}(\Delta{E}_{\uparrow}$) and $\mathbf{log}(\Delta{E}_{\downarrow}$), where
$\Delta{E}_{\uparrow}$=($E_{1\uparrow}$-$E_{2\uparrow}$) and
$\Delta{E}_{\downarrow}$=($E_{1\downarrow}$-$E_{2\downarrow}$).}
\end{figure}

To illustrate the above physical picture more clearly, with the same
parameters as Fig. 1, we plot the energy of the edge states at
$k_x=0$, namely $E_{1\uparrow}$, $E_{1\downarrow}$, $E_{2\uparrow}$
and $E_{2\downarrow}$ shown in Fig. 1(f), as a function of sample width $L_y$ in Fig. 2
(a). When $L_y$ is large, the up and down bands for spin up (
down) touche at $k_x=0$. All the edge bands are gapless and it is in
QSH regime. Decreasing the width, we can see that a measurable gap
(larger than 0.1 meV) appears for the down spin, but the up spin is
still gapless. In this QAH regime, the transport behavior is the
same as the QAH state. Decreasing further, we enter the EC regime,
where all the edge bands are gapped and the up band of spin down is
lower than the down band of spin up. In this regime, we always have
EC channels with insulating bulk. But the conductance of  EC could be completely
destroyed by disorder. When the up band of spin down touches the down
band of spin up, we get a critical point which separates the EC
regime from the NI regime [see Fig. 1(d)]. In NI regime, the system
is insulating for both bulk and edges. In Fig. 2(b), we show that,
for both spin up and down, the gap decays exponentially as a
function of width $L_y$, if $L_y$ is not too small.

\begin{figure}
\includegraphics[width=3.4in]{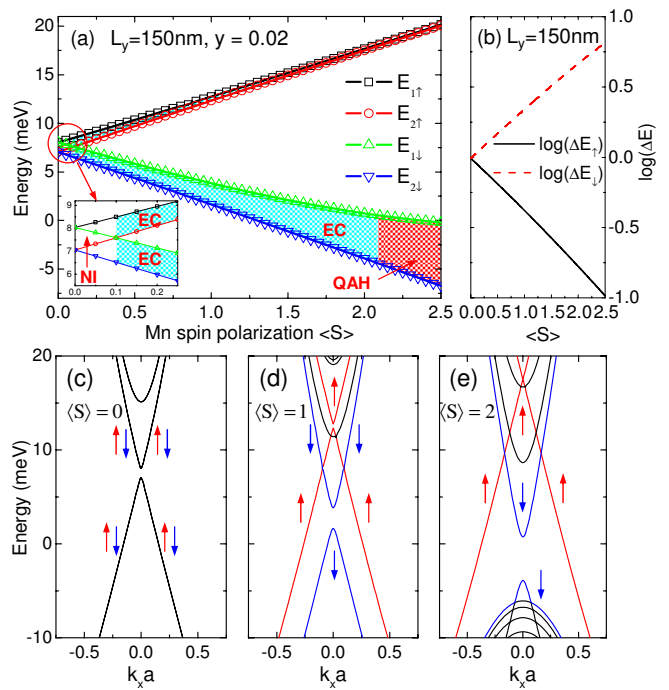}% Here is how to import EPS art
\caption{\label{fig:wide}(Color online) (a) The band energies
$E_{1\uparrow}$, $E_{2\uparrow}$, $E_{1\downarrow}$ and
$E_{2\downarrow}$ as a function of  the Mn spin polarization
$\langle{S}\rangle$  with  $L_{y}$=150nm
and the Mn doping $y=0.02$. The transport regimes are shown as well.  (b) The
corresponding energy gaps $\mathbf{log}(\Delta{E}_{\uparrow,\downarrow})$.  (c)-(e)
The energy spectra of three chosen structures with different
values of $\langle{S}\rangle$.}
\end{figure}

Given the width of the HgMnTe QW sample, different doping situation
will give different transport behaviour. As shown above, the
magnetic doping in our model is described by the spin splitting,
which is determined by the product of doping concentration $y$ and
local spin polarization $\langle{S}\rangle$. Take $L_{y}=150$nm as
example, we illustrate the variation of the transport regime when
changing the doping situation. Here, we set $y=0.02$ and change the
value of $\langle{S}\rangle$. We calculate the energy of edge bands,
as well as the gap, for both spin up and down as a function of
$\langle{S}\rangle$. The results are shown in Fig. 3(a) and (b).
With similar analysis, increasing the spin polarization, we can see
that the system will evolve from the NI regime [Fig. 3(c)] into the
EC regime [Fig. 3(d)] and finally into the QAH regime [Fig. 3(e)].
Nevertheless, the difference from the case of changing
sample width (see Fig. 2) is that,  increasing the spin
polarization, the energy gap for down spin decays exponentially,
while the gap for up spin is just opposite, as shown in Fig. 3(b).
This implies that for a narrow HgMnTe sample in the NI regime,
enhancing the Mn spin polarization is still an effective route to
realize QAH effect.

Now we study the whole diagram of the transport regime with three
tunable parameters: sample width $L_y$, spin polarization
$\langle{S}\rangle$ and Mn doping concentration $y$. In Fig. 4 (a)
and (b), the gap of edge states for up spin and down spin is plotted
in $L_{y}$-$\langle S \rangle$ plane, respectively. The results in
$L_{y}$-$y$ plan are plotted in Fig. 4(c) and (d). Figure 4 summarizes
the central result of this paper, where white lines represent
boundaries between different regimes. We assume that only the gap
larger than 0.1 meV can be detected in transport experiment, which
is used as a criteria of the gap opening. At zero $\langle S
\rangle$ (or $y=0$) limit, the problem is reduced to the finite size
effect of QSH system, and our results are well consistent with
previous work. \cite{17} An important feature is that the finite
size effect can fundamentally change the transport character. With
the same spin polarization, each transport regime has its own
permitted width range. For example, if one wants to get only one
quantized conducting channel, i.e. the QAH transport regime, the
width of sample should be in the range from 160 nm to 280 nm when
$\langle S \rangle=1$ and $y=0.02$. The permitted width range for
the transport regime is important for the device application because
that one has to choose proper device width to get the right
transport property. Meanwhile, it is also important for the
nanomaterials with QAH phase since the coupling between edge  states normally can not be ignored in these materials.
Another important conclusion is that the minimal spin polarization
(or doping concentration), required to obtain the QAH transport
regime, depends on the sample width, which is small in moderate
width region.

\begin{figure}
\includegraphics[width=3.5in]{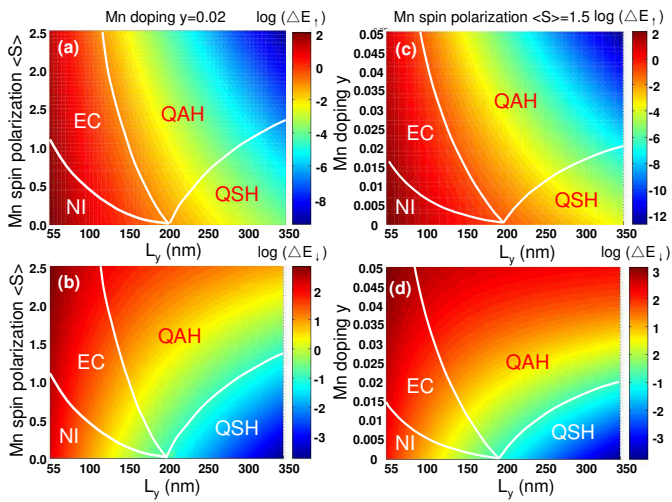}% Here is how to import EPS art
\caption{\label{fig:epsart}(Color online) Phase diagram in the
($\langle{S}\rangle$, $L_{y}$) plane for the Mn doping $y=0.02$ is
shown in the figures (a) and (b), and the phase diagram in the ($y$,
$L_{y}$) plane for the Mn spin polarization $\langle{S}\rangle=1.5$
is shown in the figures (c) and (d). In the all figures, the energy
gaps for the up spins and down spins (log($\Delta{E_{\uparrow}}$)
and log($\Delta{E_{\downarrow}}$) are used here) are plotted, where
four kinds of transport regimes, i.e., QSH regime, QAH regime, EC
regime and NI regime, are shown.}
\end{figure}

In summary, we numerically study the finite size effect in QAH
system using HgMnTe QW as an example.  We demonstrate that the
coupling of edge states in QAH system depends on both the sample
width and the magnetic doping. Thus, choosing proper sample width
and doping, we get four different transport regimes, each of which
has fundamentally different transport properties. (1) In QSH regime,
there are two quantized conducting channel near the edge which are
robust against disorder.  There is no coupling between edge states.
The charge conductance is $2e^2/h$. (2) In QAH regime, the coupling
between the edge states of down spin becomes strong enough to open
an observable gap, while the edge bands of up spin is still gapless.
The coupling makes the backscattering by disorder possible in down
spin edge conducting channel. Hence, in this regime there is only
one topological protected conducting channel left and the charge
conductance ranges from $2e^2/h$ to $e^2/h$, depending on the
disorder and position of Fermi level. (3) In EC regime, no matter
where the Fermi level is, there are always edge states near the
Fermi level. But due to the gap opening, all the edge conducting
channels can be destroyed by disorder. So the charge conductance is
in the region between $2e^2/h$ and zero, the value of which is also
determined by the disorder and the position of Fermi level. (4) In
NI regime, it is insulating in both bulk and edge. We give the whole
diagram of the transport regime in QAH system. For each regime, the
permitted width region and the corresponding doping region required
are clearly shown in the diagram. An interesting result is that the
moderate width region is favored to get the QAH transport regime.
The transport property in QAH system is very important, because it
is not only of fundamentally importance but also is the basis of
future device applications. The quantized conductance is the key
feature of the topological nontrivial phase. Meanwhile, novel
quantum electronic device will also base on the transport property.
Our results give the relation among the sample width, magnetic
doping and the transport property. It will be useful for the
analysis and design of the transport experiments in QAH systems.

%\begin{acknowledgments}
This work was supported by the National Natural Science Foundation
of China (Nos. 11274128, 11074081 and 10804034).  J.H.G acknowledges
support from the National Natural Science Foundation of China
(Grants No. 11274129). J.T.L. acknowledges from the National Natural
Science Foundation of China (Grants No. 11304107 and No. 61371015),
and the Fundamental Research Funds for the Central Universities
(HUST:2013TS032).
%\end{acknowledgments}


\end{document}